\documentclass[aps,twocolumn, preprintnumbers]{revtex4}

\usepackage{setspace}
\usepackage{graphicx}
\usepackage{subfigure}
\preprint{The following article has been submitted to Applied Physics Letters}
\begin{document}

\title{35\% magnetocurrent with spin transport through Si}

\author{Biqin Huang}
\altaffiliation{bqhuang@udel.edu}
\author{Lai Zhao}
\affiliation{Electrical and Computer Engineering Department,
University of Delaware, Newark, Delaware, 19716}

\author{Douwe J. Monsma}
\affiliation{Cambridge NanoTech Inc., Cambridge MA 02139}
\author{Ian Appelbaum}
\affiliation{ Electrical and Computer Engineering Department,
University of Delaware, Newark, Delaware, 19716}

\begin{abstract}
Efficient injection of spin-polarized electrons into the conduction band of silicon is limited by the formation of a silicide at the ferromagnetic metal (FM)/silicon interface. In the present work, this ``magnetically-dead'' silicide (where strong spin-scattering significantly reduces injected spin polarization) is eliminated by moving the FM in the spin injector from the tunnel junction base anode to the emitter cathode and away from the silicon surface. This results in over an order-of-magnitude increase in spin injection efficiency, from a previously-reported magnetocurrent ratio of $\approx$2\% to $\approx$35\% and an estimated spin polarization in Si from $\approx$1\% to at least $\approx$15\%. The injector tunnel-junction bias dependence of this spin transport signal is also measured, demonstrating the importance of low bias voltage to preserve high injected spin polarization.
\end{abstract}

\maketitle
\newpage

\begin{figure}
  \includegraphics[width=8.5cm,height=4cm]{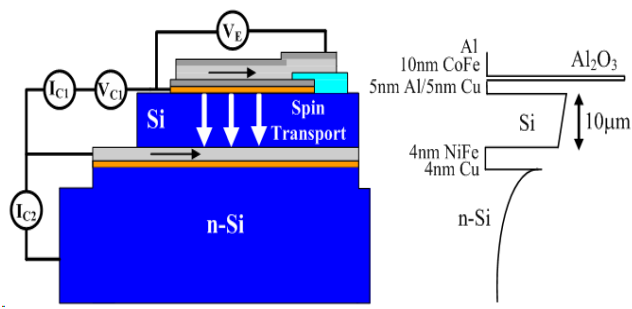}
  \caption{\label{fig:scheme}Schematic illustration of the Si spin transport device used in this work, and associated conduction band diagram (right). The vertical structure (top to bottom) is 30nm Al/10nm Co$_{84}$Fe$_{16}$/Al$_2$O$_3$/5nm Al/5nm Cu/10 $\mu m$ undoped Si/4nm Ni$_{80}$Fe$_{20}$/4nm Cu/n-Si. Spin-polarized hot electrons are injected by an emitter voltage ($V_E$) from the Co$_{84}$Fe$_{16}$ tunnel junction cathode through the normal-metal Al/Cu anode base and into the conduction band of the 10$\mu m$-thick undoped Si drift layer forming injected current $I_{C1}$. Detection on the other side is with spin-dependent hot electron transport through the Ni$_{80}$Fe$_{20}$ thin film. Our spin-transport signal is the ballistic current transported into the conduction band of the n-Si collector ($I_{C2}$).}
\end{figure}

Spin injection, manipulation and detection in semiconductors is the key to realizing the next generation of information processing devices\cite{ZUTICRMP, OHNO98}. Although magnetic semiconductors have shown some promising results\cite{Fiederling, Ohno1999, Jonker}, devices based on traditional semiconductors are often preferred. Optical techniques\cite{Kato2004,Sih2005,Gupta2001} are widely employed to study spin transport in direct band gap semiconductors, but they are inneffective for the indirect band gap semiconductor silicon (Si), which is the most dominant electronic material in the modern microelectronics industry.\cite{ZUTICPRL} Electrical techniques are preferred for device scalability, but those methods have required epitaxial ferromagnet/semiconductor growth\cite{CROWELLNATPHYS} and there are few of these systems available. We have recently solved this problem by presenting a demonstration of all-electrical spin injection, transport, precession, and detection in silicon with a device using hot-electron transport for injection and detection\cite{appelbaumnature}, which paves the way to intimately integrate the information storage properties of metallic ferromagnets with the information processing capability of Si.

\begin{figure}
    \includegraphics[width=6.25cm,height=10.5cm]{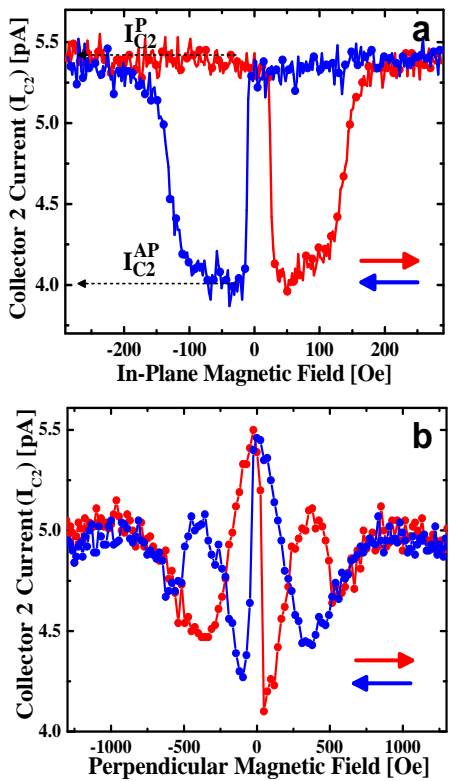}
   \caption{(a) In-plane spin-valve effect for our silicon spin transport device with emitter tunnel junction bias $V_E=$-1.6V and $V_{C1}=$0V at 85K, showing $\approx$35\% magnetocurrent. (b) Spin precession and dephasing (Hanle effect), measured by applying a perpendicular magnetic field. }
\end{figure}

Although this device finally allows study of spin-coherent electron transport in silicon, the spin-transport signal is only an approximately 2\% change in magnetocurrent, which is most likely not significant enough to be directly employed in real applications.\cite{appelbaumnature} Since hot-electron transport through a ferromagnetic metal thin film (which we used for injection and detection) results in $\approx$ 90\% spin polarization\cite{Monsma}, there must be some effect that has limited the observed spin polarization in these Si spin transport devices. Because these devices employed FM/Si interfaces at the injector and detector contacts, one obvious cause could be the ``magnetically-dead'' silicide layer formed between the silicon and ferromagnetic metals used for injector and detector\cite{VEUILLEN1987,Tsay19992,Tsay1999}. Due to strong spin-scattering between the injected electrons and the randomly-oriented magnetic moments of the metal atoms in this layer, the initial spin polarization will be suppressed, causing very low spin injection into the silicon conduction band. In this paper, we show that elimination of this FM/Si interface at the spin injector (and therefore the silicide that forms there) enables over an order-of-magnitude increase in magnetocurrent and therefore allows a much higher spin injection efficiency.

A schematic illustration of the device we use in side-view is shown in Fig. 1, together with its associated band-diagram. This device is fabricated in a way identical to the device presented in prior work\cite{appelbaumnature} except that the spin injector tunnel junction has been modified. In Ref. \cite{appelbaumnature}, the injector structure was 40nm Al/Al$_2$O$_3$/5nm Al/5nm Co$_{84}$Fe$_{16}$. Unpolarized electrons tunneled from the normal metal (NM) Al emitter cathode and were subsequently spin polarized by spin-dependent hot-electron transport through the Co$_{84}$Fe$_{16}$ base anode layer before injection over the Schottky barrier with the undoped Si drift layer. In the device used for the present work, we have placed the Co$_{84}$Fe$_{16}$ layer in the emitter cathode adjacent to the tunnel barrier oxide, so the injector structure is now 30nm Al/10nm Co$_{84}$Fe$_{16}$/Al$_2$O$_3$/5nm Al/5nm Cu. Electrons injected into the Si conduction band are already spin polarized \emph{before} hot-electron transport through the NM Al/Cu base anode. This design eliminates strong magnetic moments from adjacent FM layers in the silicide which forms at the metal/Si interface. Despite the presumably smaller initial spin polarization, this change in device design results in a higher spin polarization injected into the Si conduction band by eliminating strong spin scattering at the injector Schottky interface.

\begin{figure}
    \includegraphics[width=6.25cm,height=10.5cm]{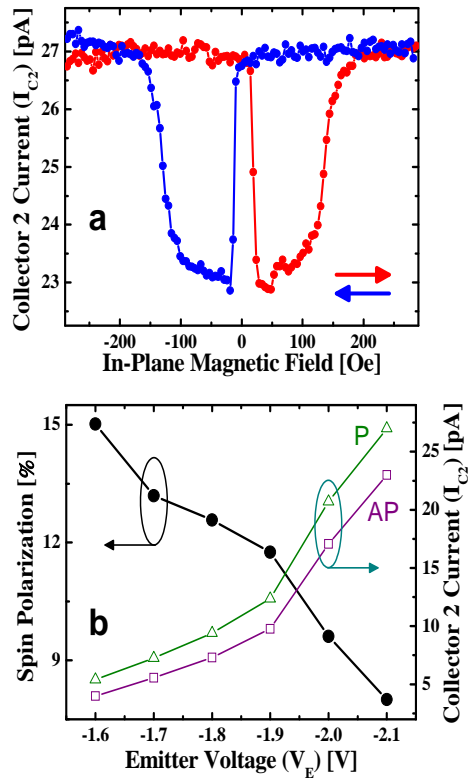}
   \caption{(a) In-plane spin valve effect for the device with $V_E=$-2.1V emitter bias and $V_{C1}=$0V at 85K, with magnetocurrent ratio reduced to $\approx$20\%. (b) $I_{C2}$ for parallel (P) and antiparallel (AP) injector/detector magnetization configuration (open symbols and right axis), and the derived lower-bound to electron spin polarization in the conduction band of Si (closed symbol and left axis).}
\end{figure}

After transport through the 10 $\mu m$-thick Si drift region, the spin polarization of conduction electrons is detected by ejecting them over a Schottky barrier and into hot-electron states in a Ni$_{80}$Fe$_{20}$ film on the other side. Because of spin-dependent scattering in this FM layer, the magnitude of the ballistic component of this hot-electron current collected by a n-Si Schottky on the other side of the film ($I_{C2}$, as shown in Fig. 1) is proportional to the spin polarization. This effect is analogous to the way an optical polarization analyzer will modulate the amount of light passing through it, depending on the relative orientation between photon linear polarization and polarization axis of the analyzer.

Fig. 2 (a) shows hysteresis of our spin-transport signal $I_{C2}$ at constant emitter bias $V_E=$-1.6V for in-plane magnetic field, where we can control the relative orientation of injected spin polarization and detection axis since the FM layers which determine these parameters have different coercive fields. The magnetocurrent ratio $(I_{C2}^P-I_{C2}^{AP})/I_{C2}^{AP}$, where the superscripts refer to parallel and anti-parallel FM injector/detector magnetization configuration, respectively, is approximately 35\%, in contrast to $\approx$2\% with our previously reported device having a FM/Si interface at the injector.\cite{appelbaumnature}

Since a spin-valve effect is a necessary but \emph{not} sufficient condition to conclude the presence of spin transport, spin precession and dephasing measurements of $I_{C2}$ in a perpendicular magnetic field\cite{JOHNSON1988,JOHNSON1985} were performed, as shown in Fig. 2 (b). Due to an in-plane component of the largely perpendicular external magnetic field, the in-plane magnetizations of the Co$_{84}$Fe$_{16}$ and Ni$_{80}$Fe$_{20}$ injector/detector FMs are parallel to each other when the measurement starts at large magnetic field, but becomes antiparallel as the field is swept through zero. At approximately 500 Oe, the parallel magnetization configuration is regained. Therefore, a complete parallel magnetization curve can be constructed from the (red) left-to-right sweep in negative field and the (blue) right-to-left sweep in positive fields. The magnetocurrent ratio seen in this measurement is consistent with the $\approx$35\% effect seen with in-plane spin-valve measurements above.

Although these results clearly demonstrate highly efficient spin injection into silicon by the large relative fluctuation in magnetic fields causing spin-valve effect and coherent precession, the absolute magnitude of I$_{C2}$ is small. To increase I$_{C2}$, we increase the bias on the tunnel junction in order to increase the electron injection current $I_{C1}$, which drives $I_{C2}$. Fig. 3 (a) shows the in-plane spin-valve effect with $V_E=$-2.1 V, where $I_{C2}$ is significantly increased. However, the magnetocurrent ratio ($\approx$20\%) is smaller than in Fig. 2 (a) where $V_E=$-1.6V (but still roughly an order of magnitude higher than previously reported)\cite{appelbaumnature}. 

Fig. 3 (b) shows I$_{C2}$ for both parallel (P) and antiparallel (AP) magnetization configuration for $V_E$ at intermediate values (open symbols and right axis). Clearly, $I_{C2}$ monotonically increases with the increase of tunnel junction emitter bias for both configurations as expected, although as noted previously the magnetocurrent ratio decreases.

Because we are interested in spin transport in Si for applications to real spintronic devices, we calculate the observed conduction electron spin polarization $\mathcal{P}=(I_{C2}^P-I_{C2}^{AP})/(I_{C2}^{P}+I_{C2}^{AP})$, shown in closed symbols and on the left axis in Fig. 3 (b). Apparently, the spin polarization decreases with increase of the emitter bias, from approximately 15\% to approximately 8\% across this bias range. This may be due to the bias dependence of spin injection from a tunnel junction\cite{Moodera1995, SergioDouwe2005}. In comparison, spin polarization deduced from prior work\cite{appelbaumnature} is only approximately 1\%. However, since our hot-electron spin detector is not a perfect spin filter, these spin polarization values are merely lower bounds. For instance, with detection efficiency $0<\mathcal{E}<1$, the actual electron spin polarization is $\mathcal{P}/(2\mathcal{E}-1)$. Assuming $\mathcal{E}=0.9$, the maximum spin polarization we deduce (at $V_E=$-1.6V) is as high as $\approx$19\%.  

In summary, we have presented an all-electrical device with highly efficient spin injection into silicon. This improvement of over an order of magnitude compared to previous work was enabled by spin polarization at the emitter/tunnel-junction-barrier interface, which eliminates the possibility of silicide formation with a FM at the spin injector metal/Si Schottky. It will improve our ability to study spin transport in silicon and is particularly important for silicon spintronic applications. We predict that even higher spin injection efficiency may be obtained with a FM/Oxide/FM/NM spin injector.

This work was supported in part by DARPA/MTO and the Office of Naval Research.


\begin{thebibliography}{10}


\bibitem{ZUTICRMP}
I. Zutic, J. Fabian, S. Das Sarma, Rev. Mod. Phys. {\bf{76}}, 323 (2004).

\bibitem{OHNO98}
H. Ohno, Science {\bf{281}}, 951 (1998).



\bibitem{Fiederling}
R. Fiederling, M. Keim, G. Reuscher, W. Ossau, G. Schmidt, A. Waag and L. W. Molenkamp, Nature {\bf{402}}, 787 (1999).

\bibitem{Ohno1999}
Y. Ohno, D. K. Young, B. Beschoten, F. Matsukura, H. Ohno, and D. D.
Awschalom, Nature {\bf{402}}, 790 (1999).

\bibitem{Jonker}
B. T. Jonker, Y. D. Park, B. R. Bennett, H. D. Cheong,
G. Kioseoglou, and A. Petrou, Phys. Rev. B {\bf{62}}, 8183 (2000). 

\bibitem{Kato2004}
Y. K. Kato, R.C. Myers, A.C. Gossard, and D. D. Awschalom, 
Science {\bf{306}}, 1910 (2004).

\bibitem{Sih2005}
V. Sih, R. C. Myers, Y. K. Kato, W. H. Lau, A. C. Gossard, and D. D.
Awschalom, Nature Phys. {\bf{1}}, 31 (2005).

\bibitem{Gupta2001}
J. A. Gupta, R. Knobel, N. Samarth, and D. D. Awschalom, Science {\bf{292}}, 2458 (2001).

\bibitem{ZUTICPRL}
I. Zutic, J. Fabian, and S. Erwin, Phys. Rev. Lett., {\bf{97}}, 026602 (2006).

\bibitem{CROWELLNATPHYS}
X. Lou, C. Adelmann, S. A. Crooker, E. S. Garlid, J. Zhang, S. M. Reddy, S. D. Flexner, C. J. Palmstrom, and P. A. Crowell, Nature Phys. {\bf{3}}, 197 (2007)


\bibitem{appelbaumnature}
I. Appelbaum, B. Huang, and D. Monsma, Electronic measurement and
control of spin transport in silicon (2007), URL http://lanl.arxiv.org/abs/cond-mat/0703025.

\bibitem{Monsma}
D.J. Monsma, R. Vlutters, and J.C. Lodder, Science {\bf{281}}, 407 (1998).

\bibitem{VEUILLEN1987}
J. Y. Veuillen, J. Derrien, P. A. Badoz, E. Rosencher, and C. Danterroches, Appl. Phys. Lett. {\bf{51}}, 1448 (1987).

\bibitem{Tsay19992}
J. S. Tsay and Y. D. Yao, Appl. Phys. Lett. {\bf{74}}, 1311 (1999).

\bibitem{Tsay1999}
J. S. Tsay, C. S. Yang, Y. Liou, and Y. D. Yao, J. Appl. Phys. {\bf{85}}, 4967 (1999).

\bibitem{JOHNSON1988}
M. Johnson and R. H. Silsbee, Phys. Rev. B  {\bf{37}}, 5326 (1988).

\bibitem{JOHNSON1985}
M. Johnson and R. H. Silsbee, Phys. Rev. Lett. {\bf{55}}, 1790 (1985).

\bibitem{Moodera1995}
J. S. Moodera, L. R. Kinder, T. M. Wong, and R. Meservey, Phys. Rev. Lett. {\bf{74}}, 3273 (1995).

\bibitem{SergioDouwe2005}
S. Valenzuela, D.J. Monsma, C.M. Marcus, V. Narayanamurti, and M. Tinkham, Phys. Rev. Lett. {\bf{94}}, 196601 (2005).

\end{thebibliography}
\end{document}